\documentclass[pra,showpacs,superscriptaddress]{revtex4}
\usepackage{graphicx}
\usepackage{dcolumn}
\usepackage{bm}
\usepackage{amsmath}
\usepackage{amssymb,amsbsy}
\usepackage{color,ulem} 
\newcommand{\beq}{\begin{equation}}
\newcommand{\eeq}{\end{equation}}
\newcommand{\bea}{\begin{eqnarray}}
\newcommand{\eea}{\end{eqnarray}}
\newcommand{\bec}{\begin{center}}
\newcommand{\enc}{\end{center}}
\newcommand{\bfr}{\begin{flushright}}
\newcommand{\efr}{\end{flushright}}
\newcommand{\la}{\langle}
\newcommand{\ra}{\rangle}

\newcommand{\alp}{\alpha}

\newcommand{\om}{\omega}
\newcommand{\kap}{\kappa}
\newcommand{\gam}{\gamma}
\newcommand{\sig}{\sigma}

\newcommand{\Om}{\Omega}

\newcommand{\ta}{\widetilde{a}}
\newcommand{\tb}{\widetilde{b}}

\newcommand{\tom}{\widetilde{\omega}}

\newcommand{\tkap}{\widetilde{\kappa}}
\newcommand{\tone}{\widetilde{1}}
\newcommand{\ttwo}{\widetilde{2}}
\newcommand{\tthree}{\widetilde{3}}
\newcommand{\tfour}{\widetilde{4}}

\newcommand{\cH}{{\cal H}}

\begin{document}
\title{Theory of microwave single-photon detection 
using an impedance-matched $\Lambda$ system}
\author{Kazuki Koshino}
\affiliation{College of Liberal Arts and Sciences, Tokyo Medical and Dental
University, Ichikawa, Chiba 272-0827, Japan}
\author{Kunihiro Inomata}
\affiliation{RIKEN Center for Emergent Matter Science (CEMS), 2-1 Hirosawa, Wako, 
Saitama 351-0198, Japan}
\author{Zhirong Lin}
\affiliation{RIKEN Center for Emergent Matter Science (CEMS), 2-1 Hirosawa, Wako, 
Saitama 351-0198, Japan}
\affiliation{Beijing Computational Science Research Center, Beijing 100084, China}
\author{Yasunobu Nakamura}
\affiliation{RIKEN Center for Emergent Matter Science (CEMS), 2-1 Hirosawa, Wako, 
Saitama 351-0198, Japan}
\affiliation{Research Center for Advanced Science and Technology (RCAST), 
The University of Tokyo, Meguro-ku, Tokyo 153-8904, Japan}
\author{Tsuyoshi Yamamoto}
\affiliation{RIKEN Center for Emergent Matter Science (CEMS), 2-1 Hirosawa, Wako, 
Saitama 351-0198, Japan}
\affiliation{Smart Energy Research Laboratories, NEC Corporation,  
Tsukuba, Ibaraki 305-8501, Japan}
\date{\today}
\begin{abstract}
By properly driving a qubit-resonator system in the strong dispersive regime, 
we implement an ^^ ^^ impedance-matched'' $\Lambda$ system in the dressed states,
where a resonant single photon deterministically induces 
a Raman transition and excites the qubit. 
Combining this effect and a fast dispersive readout of the qubit, 
we realize a detector of itinerant microwave photons. 
We theoretically analyze the single-photon response of the $\Lambda$ 
system and evaluate its performance as a detector. 
We achieve a high detection efficiency close to unity 
without relying on precise temporal control of the input pulse shape
and under a conservative estimate of the system parameters.
The detector can also be reset quickly by applying microwave pulses,
which allows a short dead time and a high repetition rate. 
\end{abstract}
\maketitle

\section{Introduction}
Extensive efforts have been made in a variety of physical systems 
to realize strong coupling between a single quantum emitter 
and a one-dimensional photon field~\cite{1d1,1d2,1d3,1d4,1d5,1d6,1d7,1d8,1d9,1d10}. 
In such one-dimensional optical systems, 
interaction between an emitter and a photon is enhanced drastically 
due to the destructive interference between 
the incident field and the radiation from the emitter. 
This opens the possibility for deterministic control 
of quantum systems by individual photons. 

In particular, in a $\Lambda$-type three-level system 
that has identical radiative decay rates from the top level 
and is coupled to a semi-infinite one-dimensional field, 
a resonant single photon deterministically induces a Raman transition 
in the $\Lambda$ system and switches its electronic state~\cite{Lam1,Lam2,Lam3,Lam4}. 
Recently, we implemented such a $\Lambda$ system by utilizing the dressed states 
of a driven circuit-quantum-electrodynamics (circuit-QED) system. 
Applying continuous microwave to the $\Lambda$ system, 
we confirmed that the field amplitude vanishes completely 
upon its reflection (perfect absorption by ^^ ^^ impedance matching''), 
and that input photons are down-converted 
by the $\Lambda$ system~\cite{IML1,IML2,IML3}. 
This indicates highly efficient switching of the $\Lambda$ system 
induced by individual microwave photons. 
Deterministic switching of a $\Lambda$ system 
has also been demonstrated recently in the visible-light domain 
by using a spherical cavity and an atom~\cite{Dayan1,Dayan2}.

Exploiting the large transition dipole of superconducting qubits, 
circuit QED enables various microwave quantum-optical phenomena 
that have not been reached by quantum optics in the visible domain:
The examples include direct and strong coupling of qubits 
to itinerant waveguide photons~\cite{wqed1,wqed2,wqed3}
and various effects in the strong-dispersive regime 
of the qubit-cavity interaction~\cite{sd1,sd2,sd3}. 
However, an apparent shortcoming of microwave quantum optics has been 
the lack of efficient single photon detectors: The main difficulty is in 
the energy scale of single microwave photons orders of magnitude smaller 
than that of visible or infrared photons.

In recent proposals and an experiment,
a current-biased Josephson junction was used 
as a microwave photon detector~\cite{solano1,solano2,McDer}:
a metastable two-level system formed in the tilted washboard potential
resonantly absorbs an incoming photon and 
switches into the voltage state of the junction to make a ^^ ^^ click''.
While the successful operation was demonstrated~\cite{McDer},
the dark count and the dead time after the detection are remaining issues.
In this study, we theoretically analyze the single-photon response of 
an impedance-matched $\Lambda$ system 
realized in a driven qubit-resonator system
and demonstrates its excellent performance 
as a single photon detector in the microwave domain. 
The detector has a low dark count rate
and attains a high detection efficiency close to unity.
Moreover, the system is switched to a discrete 
and definite quantum state upon detection,
which enables us to reset the system quickly 
and therefore to shorten the dead time of the detector. 
In comparison with the works that demonstrate 
deterministic capture of propagating microwave pulses 
in a harmonic oscillator mode~\cite{catch1,catch2,catch3,catch4},
our scheme based on a $\Lambda$ system has a merit that
it is free from precise temporal control 
of the input pulse shape and/or the system parameters: 
high switching efficiency close to unity can be attained
as long as the input photon has a narrower bandwidth
than the linewidth of the $\Lambda$ system.

The rest of this paper is organized as follows.
In Sec.~\ref{sec:sys}, we theoretically describe 
the driven qubit-resonator system 
and derive basic equations for the analysis.
In Sec.~\ref{sec:mode}, we explain two distinct functionalities of the device. 
In the ^^ ^^ $\Lambda$'' mode, which is realized under a proper qubit drive,
a single incident photon excites the qubit deterministically.
In the ^^ ^^ I'' mode, which is realized when the qubit drive is off,
the dispersive readout of the qubit state is possible.
In Sec.~\ref{sec:scheme}, we present our single-photon detection scheme,  
which is composed of three stages: capture, readout, and reset. 
In Sec.~\ref{sec:capture}, we discuss 
the capture stage and show that a high detection efficiency 
close to unity is achievable with practical parameter values. 
In Sec.~\ref{sec:reset}, we discuss the reset stage and show that 
the present system can be initialized within a few hundred nanoseconds.
Section~\ref{sec:summary} summarizes this study.

\begin{figure}[t]
\includegraphics[scale=1.0]{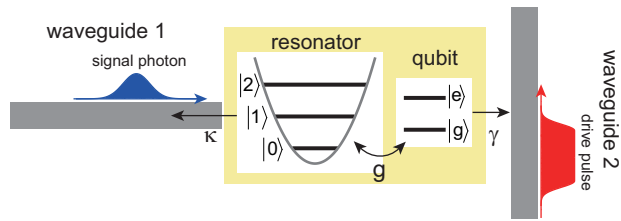}
\caption{Schematic of the setup.
A superconducting qubit (two-level atom) is coupled dispersively to a resonator.
The resonator is coupled to a semi-infinite waveguide (WG1),
through which a signal photon is input.
The qubit is coupled to another waveguide (WG2),
through which a drive pulse is applied.}
\label{fig:sch}
\end{figure}

\section{System}\label{sec:sys}
Here, we theoretically describe the device considered in this study~\cite{dev1,dev2}.
Its schematic is shown in Fig.~\ref{fig:sch}.
A superconducting qubit, which can be regarded as a two-level atom,
is dispersively coupled to a transmission line resonator.
The resonator is further coupled to a semi-infinite waveguide (WG1),
through which a signal photon pulse to be detected is input. 
Our objective is to determine
whether the signal pulse contains a photon or not. 
We also use this port for readout of the qubit and for resetting the system. 
Through another waveguide (WG2), we apply drive pulses to the qubit 
in order to engineer the dressed states of the qubit-resonator system.

\subsection{Hamiltonian}
Setting $\hbar=v=1$, 
where $v$ is the microwave velocity in the waveguides,
the Hamiltonian of the overall system is written as
\bea
\cH &=& \cH_{\rm sys}+\cH_{\rm damp},
\label{eq:H}
\\
\cH_{\rm sys} &=& 
\om_r a^{\dag}a \sig \sig^{\dag} +
\left[\om_q+(\om_r-2\chi)a^{\dag}a \right] \sig^{\dag}\sig,
\label{eq:Hsys}
\\
\cH_{\rm damp} &=& 
\int dk \left[ k a_k^{\dag}a_k + \sqrt{\kap^{\prime}/2\pi}(a^{\dag}a_k + a_k^{\dag}a)\right]
+\int dk \left[ k b_k^{\dag}b_k + \sqrt{\gam^{\prime}/2\pi}
(\sig^{\dag}b_k + b_k^{\dag}\sig)\right],
\label{eq:Hdamp}
\eea
where $a$ ($\sig$) is the annihilation operator for the resonator (qubit)
and $a_k$ ($b_k$) is the annihilation operator for microwave photon 
propagating in WG1 (WG2) with wave number $k$. 
$\om_r$ ($\om_q$) is the resonance frequency of the resonator (qubit),
$\chi$ is the dispersive frequency shift,
and $\kap^{\prime}$ ($\gam^{\prime}$) is the radiative decay rate 
of the resonator (qubit) into WG1 (WG2).
The parameter values are listed in Table~\ref{tab:1}.

Three comments are in order regarding this Hamiltonian. 
(i)~The qubit-resonator system is described by the Jaynes-Cummings Hamiltonian, 
$\cH_{\rm sys}=\bar{\om}_r a^{\dag}a + \bar{\om}_q \sig^{\dag}\sig + g(a^{\dag}\sig+\sig^{\dag}a)$,
where $\bar{\om}_r$ and $\bar{\om}_q$ are the bare frequencies of the resonator and qubit
and $g$ is their coupling. 
In the dispersive regime ($|\bar{\om}_r - \bar{\om}_q| \gg g$),
$\cH_{\rm sys}$ is recast into a diagonal form of Eq.~(\ref{eq:Hsys}). 
The renormalized frequencies are given by
$\om_r=\bar{\om}_r+\chi$ and 
$\om_q=\bar{\om}_q-\chi$, where $\chi=g^2/(\bar{\om}_r-\bar{\om}_q)$. 
(ii)~Although omitted for simplicity,
the resonator and qubit are subject to nonradiative decay to the environment. 
We denote the total decay rate of the resonator (qubit) by $\kap$ ($\gam$).
(iii)~In practice, we switch to the rotating frame 
to remove the natural phase factors. 
We subtract $\cH_0=\om_s a^{\dag}a + \om_d \sig^{\dag}\sig$
from $\cH_{\rm sys}$ of Eq.~(\ref{eq:Hsys}), 
where $\om_s$ ($\om_d$) is the central frequency 
of the signal photon (drive pulse).

\begin{table}[t]
\begin{center}
    \begin{tabular}{|l|l|} 
       \hline
       $\om_q$ & $2\pi \times  5$~GHz \\
       $\om_r$ & $2\pi \times 10$~GHz \\ 
       $\chi$   & $2\pi \times 40$~MHz \\ 
       \hline
       $\kap$    & $2\pi \times 20$~MHz \\
       $\kap^{\prime}$  & $2\pi \times 20$~MHz \\
       $\gam$   & $2\pi \times 0.1$~MHz \\
       $\gam^{\prime}$ & $2\pi \times 0.1$~kHz \\ 
       \hline
       $\om_d$ & $\om_q-2\pi\times 70$~MHz \\
       \hline
    \end{tabular}
\caption{Parameter values for $\om_q$ (qubit frequency), 
$\om_r$ (resonator frequency), $\chi$ (dispersive shift), 
$\kap$ (total decay rate of the resonator), 
$\kap^{\prime}$ (radiative decay rate of the resonator to WG1), 
$\gam$ (total decay rate of the qubit), and 
$\gam^{\prime}$ (radiative decay rate of the qubit to WG2),
and $\om_d$ (qubit drive frequency). 
\label{tab:1}
}
\end{center}
\end{table}

\subsection{Heisenberg equations}
We introduce the real-space representation of the field operator for WG1 by
\beq
\ta_r =\frac{1}{\sqrt{2\pi}}\int dk e^{ikr} a_k.
\eeq
In this representation, 
the $r<0$ ($r>0$) region corresponds to the incoming (outgoing) field.
The field operator $\tb_r$ for WG2 is defined similarly.
From $\cH_{\rm damp}$ of Eq.~(\ref{eq:Hdamp}), 
we can derive the following input-output relations: 
\bea
\ta_r(t) &=& \ta_{r-t}(0)-i\sqrt{\kap^{\prime}}a(t-r)\theta(r)\theta(t-r),
\label{eq:art}
\\
\tb_r(t) &=& \tb_{r-t}(0)-i\sqrt{\gam^{\prime}}\sig(t-r)\theta(r)\theta(t-r).
\label{eq:brt}
\eea
The conventional input and output field operators are defined at $r=\pm 0$ by
$a_{\rm in}(t)=\ta_{-0}(t)$ and $a_{\rm out}(t)=\ta_{+0}(t)$.
$b_{\rm in}(t)$ and $b_{\rm out}(t)$ are defined similarly.
From Eqs.~(\ref{eq:H}), (\ref{eq:art}) and (\ref{eq:brt}),
the Heisenberg equation for any system operator $S$
(composed of $\sig$, $a$ and their conjugates) is given by
\bea
\frac{d}{dt}S &=& i[\cH_{\rm sys}, S]
+\frac{\kap}{2}(2a^{\dag}Sa-a^{\dag}aS-Sa^{\dag}a)
+\frac{\gam}{2}(2\sig^{\dag}S\sig-\sig^{\dag}\sig S-S\sig^{\dag}\sig)
\nonumber
\\
&+& i\sqrt{\kap^{\prime}}a_{\rm in}^{\dag}(t)[a,S]
+i\sqrt{\kap^{\prime}} [a^{\dag},S] a_{\rm in}(t)
+i\sqrt{\gam^{\prime}}b_{\rm in}^{\dag}(t)[\sig,S]
+i\sqrt{\gam^{\prime}} [\sig^{\dag},S] b_{\rm in}(t). 
\label{eq:Hei}
\eea

\subsection{Microwave response}
In our setup, a signal photon is input through WG1 and 
a classical drive pulse is applied through WG2.
We denote the wavefunction of the signal photon by $f_s(t)$
and the amplitude of the drive pulse by $f_d(t)$. 
Note that $f_s(t)$ is normalized as $\int dt |f_s(t)|^2=1$. 
Extension to the two-photon input through WG1 
is straightforward (see Appendix~\ref{app:A}).

Analysis of the microwave response to a single photon can be simplified by
replacing the single-photon state $|1\ra$ with a coherent state $|\alp\ra$,
performing perturbation calculation with respect to $\alp$,
and picking up the relevant terms afterwards~\cite{KKPRL2007}.
Therefore, we investigate a situation
in which two classical pulses, $\alp f_s(t)$ and $f_d(t)$, 
are applied through WG1 and WG2, respectively.
Since a classical pulse (coherent state) is 
an eigenstate of an input field operator, 
we can rigorously replace 
$a_{\rm in}(t)$ and $b_{\rm in}(t)$ in Eq.~(\ref{eq:Hei})
with $\alp f_s(t)$ and $f_d(t)$, respectively.
Then, the expectation value of an operator $S$ evolves as
\bea
\frac{d}{dt}\la S \ra_c &=& i\la [\cH_{\rm sys}, S]\ra_c 
+\kap(\la a^{\dag}Sa \ra_c - \la a^{\dag}aS\ra_c/2-\la Sa^{\dag}a\ra_c/2)
+\gam(\la\sig^{\dag}S\sig\ra_c-\la\sig^{\dag}\sig S\ra_c/2-\la S\sig^{\dag}\sig\ra_c/2)
\nonumber
\\
&+& i\sqrt{\kap^{\prime}}\alp^* f^*_t(t)\la[a,S]\ra_c 
+i \sqrt{\kap^{\prime}}\alp f_s(t) \la[a^{\dag},S]\ra_c 
+i\sqrt{\gam^{\prime}}f^*_d(t)\la[\sig,S]\ra_c
+i\sqrt{\gam^{\prime}} f_d(t)\la[\sig^{\dag},S]\ra_c. 
\eea
We expand $\la S \ra_c$ as 
$\la S \ra_c = \sum_{m,n=0}^{\infty} (\alp^*)^m \alp^n \la S \ra_c^{mn}$.
Then, $\la S \ra_c^{mn}$ evolves as
\bea
\frac{d}{dt}\la S \ra_c^{mn} &=& i\la [\cH_{\rm sys}, S]\ra_c^{mn} 
+ \kap(\la a^{\dag}Sa \ra_c^{mn} - \la a^{\dag}aS\ra_c^{mn}/2-\la Sa^{\dag}a\ra_c^{mn}/2)
+ \gam(\la\sig^{\dag}S\sig\ra_c^{mn}-\la\sig^{\dag}\sig S\ra_c^{mn}/2
- \la S\sig^{\dag}\sig\ra_c^{mn}/2)
\nonumber
\\
&+& i\sqrt{\kap^{\prime}} f^*_t(t)\la[a,S]\ra_c^{(m-1)n}
+i \sqrt{\kap^{\prime}} f_s(t) \la[a^{\dag},S]\ra_c^{m(n-1)} 
+ i\sqrt{\gam^{\prime}}f^*_d(t)\la[\sig,S]\ra_c^{mn}
+ i\sqrt{\gam^{\prime}} f_d(t)\la[\sig^{\dag},S]\ra_c^{mn}. 
\eea
with the convention that $\la S \ra_c^{mn}=0$
for negative $m$ or $n$.

Our objective is to evaluate the expectation value 
for a single-photon input, $\la S \ra_s$.
As discussed in Appendix~\ref{app:A}, it is given by
\beq
\la S \ra_s = \la S \ra_c^{00} + \la S \ra_c^{11}.
\label{eq:trans}
\eeq
Therefore, we solve the simultaneous differential equations for 
$\la S \ra_c^{00}$, $\la S \ra_c^{01}$, $\la S \ra_c^{10}$
and $\la S \ra_c^{11}$ to evaluate $\la S \ra_s$.

\section{I and $\Lambda$ modes}\label{sec:mode}
\begin{figure}[t]
\includegraphics[scale=1.0]{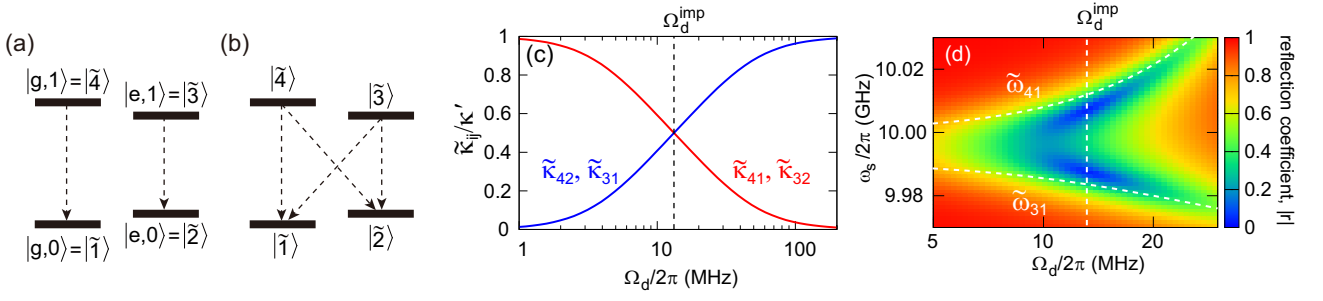}
\caption{
(a)~I mode of the qubit-resonator system. 
This is attained when the qubit drive is off. 
Arrows indicate the directions of radiative decays. 
(b)~$\Lambda$ mode of the qubit-resonator system. 
The four decay rates are identical. 
This is attained when a proper drive field is applied to the qubit. 
(c)~Radiative decay rates $\tkap_{31}$, 
$\tkap_{32}$, $\tkap_{41}$ and $\tkap_{42}$
as functions of the drive amplitude. 
The four rates become identical at $\Om_d^{\rm imp}/2\pi=13.2$~MHz.
(d)~Reflection coefficient $|r|$ of a weak 
signal field applied through WG1.
White dashed lines show $\tom_{31}$ and $\tom_{41}$. 
Impedance matching ($|r|\simeq 0$) occurs at 
$(\Om_d, \om_p) \simeq (\Om_d^{\rm imp}, \tom_{31})$ 
and $(\Om_d^{\rm imp}, \tom_{41})$.
}
\label{fig:cr}
\end{figure}

As discussed in Refs.~\cite{IML1,IML2}, 
for a drive field with properly chosen power and frequency,
the dressed states of a qubit-resonator system constitutes 
an ^^ ^^ impedance-matched'' $\Lambda$ system, 
which absorbs an input photon with a high efficiency close to unity. 
To observe this, we first consider a case 
with a continuous drive in this section. 
The drive field $f_d(t)$ is given by
\beq
f_d(t)=\frac{\Om_d}{\sqrt{\gam^{\prime}}} e^{-i\om_d t},
\eeq
where $\om_d$ is the drive frequency
and $\Om_d$ is the drive amplitude expressed in terms of the Rabi frequency. 
In the frame rotating at $\om_d$, 
the Hamiltonian for the driven qubit-resonator system is written as
\bea
\cH_{\rm sys+dr} &=& 
\om_r a^{\dag}a \sig \sig^{\dag} +
\left[(\om_q-\om_d)+(\om_r-2\chi)a^{\dag}a \right] \sig^{\dag}\sig
+ \Om_d(\sig^{\dag}+\sig).
\label{eq:sysdr}
\eea
We label the state vectors of the system with $|q,n\ra$, 
where $q(=g,e)$ denotes the qubit state
and $n(=0,1,\cdots)$ denotes the resonator photon number.
Throughout this study, only the lowest four states
($|g,0\ra$, $|e,0\ra$, $|g,1\ra$ and $|e,1\ra$) are relevant.
When the drive is off ($\Om_d=0$),
their energies are respectively given by
$\om_{|g,0\ra}=0$, $\om_{|e,0\ra}=\om_q-\om_d$, 
$\om_{|g,1\ra}=\om_r$ and $\om_{|e,1\ra}=\om_q-\om_d+\om_r-2\chi$. 
We choose the drive frequency $\om_d$ to satisfy
$\om_q-2\chi<\om_d<\om_q$. 
Then, the level structure becomes nested, i.e.,
$\om_{|g,0\ra}<\om_{|e,0\ra}<\om_{|e,1\ra}<\om_{|g,1\ra}$
[Fig.~\ref{fig:cr}(a)]. 
When the drive is on ($\Om_d>0$),
$|g,0\ra$ and $|e,0\ra$ ($|g,1\ra$ and $|e,1\ra$) 
are mixed to form the dressed states 
$|\tone\ra$ and $|\ttwo\ra$ ($|\tthree\ra$ and $|\tfour\ra$),
where the dressed states are labeled from the lowest in energy
[Fig.~\ref{fig:cr}(b)].
The radiative decay rate from $|\widetilde{i}\ra$ to $|\widetilde{j}\ra$ 
emitting a photon into WG1 is given by 
$\tkap_{ij}=\kap^{\prime} |\la\widetilde{i}|a^{\dag}|\widetilde{j}\ra|^2$.

Figure~\ref{fig:cr}(c) plots $\tkap_{ij}$ as functions of the drive amplitude. 
The drive frequency $\om_d$ is fixed at the value 
in Table~\ref{tab:1} throughout this study.
When the drive is off, 
$\tkap_{32} = \tkap_{41} = \kap^{\prime}$
and $\tkap_{31} = \tkap_{42} = 0$.
Namely, the radiative decay occurs only in one direction
as depicted in Fig.~\ref{fig:cr}(a). 
We refer to this as the ^^ ^^ I'' mode of the qubit-resonator system.
In this mode, the microwave transition occurs conserving the qubit state,
and the transition frequency differs by $2\chi$ depending on the qubit state.
We can use this mode for the dispersive readout of the qubit state.

In contrast, at a proper drive power 
[$\Om_d^{\rm imp}$ in Fig.~\ref{fig:cr}(c)],
the four decay rates become identical, i.e., 
$\tkap_{31} = \tkap_{32} = \tkap_{41} = \tkap_{42} = \kap'/2$. 
Then, the three levels $|\tone\ra$, $|\ttwo\ra$ and $|\widetilde{u}\ra$ ($u=3$ or 4)
function as an impedance-matched $\Lambda$ system:
a single photon resonant to the $|\tone\ra \to |\widetilde{u}\ra$ transition
deterministically induces the Raman transition 
of $|\tone\ra \to|\widetilde{u}\ra \to |\ttwo\ra$
and switches the electronic state.
We refer to this as the ^^ ^^ $\Lambda$'' mode  
of the qubit-resonator system [Fig.~\ref{fig:cr}(b)].
The underlying physical mechanism for this phenomenon 
is the destructive interference between 
the input and elastically scattered photons~\cite{IML1}.
Figure~\ref{fig:cr}(d) shows the reflection coefficient $|r|$
of a weak continuous signal field applied through WG1, 
as a function of the drive amplitude $\Omega_d$ and the signal frequency $\om_s$.
We observe two impedance-matching spots ($|r|\simeq 0$)
at $(\Om_d, \om_s) \simeq (\Om_d^{\rm imp}, \tom_{31})$ 
and $(\Om_d^{\rm imp}, \tom_{41})$,
where $\tom_{ij}$ is the transition frequency 
between dressed states $|\widetilde{i}\ra$ and $|\widetilde{j}\ra$.
We have two comments regarding Fig.~\ref{fig:cr}(d). 
(i)~We observe that $\om_s$ of the upper (lower) spot 
deviates slightly from $\tom_{41}$ ($\tom_{31}$). 
This is due to the small energy difference between 
levels $|\tthree\ra$ and $|\tfour\ra$
that is comparable to the linewidth of the $\Lambda$ system ($\sim \kap'/2$). 
In this case, weak elastic scattering induced by level 
$|\tthree\ra$ ($|\tfour\ra$) is not negligible.
(ii)~In our former experiment, $\Om_d$ of the two spots differed considerably~\cite{IML3}.
That was because we used a relatively strong signal field 
to improve the signal to noise ratio. 
Here, we consider the weak-field limit to discuss the single-photon response.

\section{Detection scheme}\label{sec:scheme}
As we observed in the previous section, 
the present qubit-resonator system has two distinct functionalities. 
In the $\Lambda$ mode, the system captures a single photon 
nearly deterministically and makes a transition to the qubit excited state. 
In the I mode, the system functions as an oscillator 
whose resonance frequency depends on the qubit state. 
We can read out the qubit state efficiently with a single probe pulse. 
These two modes can be switched adiabatically 
by a temporally smooth qubit drive pulse.

In the single-photon detection considered here, 
it is assumed that the arrival time and the length 
of the signal pulse are known in advance:
we determine whether the pulse contains 
a single photon or not by this measurement.
Our single-photon detection proceeds in the following three steps.
(i)~Initially, the qubit-resonator system is in its ground state $|g,0\ra$
and a signal pulse is input through WG1. 
Through WG2, we apply a drive pulse that covers the signal pulse in time, 
and switch the system to the $\Lambda$ mode during this period. 
At the end of this stage, the photon number in the signal pulse (0 or 1)
is mapped onto the qubit state ($g$ or $e$).
We refer to this as the ^^ ^^ capture'' stage
and discuss in detail in Sec.~\ref{sec:capture}.
(ii)~Next, we perform dispersive readout of the qubit. 
In order to prevent the $|\ttwo\ra \to |\widetilde{u}\ra \to |\tone\ra$ 
transition ($u=3$ or $4$) induced by the probe, 
we turn off the qubit drive and keep the system
in the I mode during this process. 
We apply a classical probe pulse to the resonator 
and measure the phase shift of the reflected pulse.
We refer to this as the ^^ ^^ readout'' stage.
We do not touch this stage in this paper,
since dispersive readout of the qubit has been discussed in prior works:
High-fidelity single-shot readout has been achieved with 
the Josephson parametric amplifier~\cite{read2,read3} 
and the parametric phase-locked oscillator~\cite{read4}. 
(iii)~The qubit-resonator system is initialized 
(i.e., decays to its ground state $|g,0\ra$)
automatically by the natural qubit decay. 
However, in order to shorten the dead time of the detector,
we artificially reset the system by applying microwave pulses.
For this purpose, we use the inverse process of stage~(i):
We apply a drive pulse through WG2 to form a $\Lambda$ system
and a reset pulse through WG1.
Note that the reset pulse can be a classical and strong one
in contrast with the capture stage.
We refer to this as the ^^ ^^ reset'' stage
and discuss in detail in Sec.~\ref{sec:reset}.

\begin{figure}[t]
\includegraphics[scale=1.0]{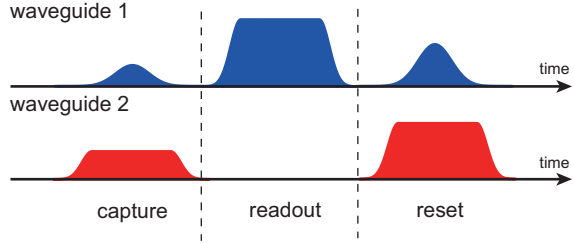}
\caption{
Pulse sequence for single-photon detection.
(i)~Capture stage. We simultaneously input a signal photon through WG1
and a classical drive pulse through WG2.
(ii)~Readout stage. We apply a classical readout pulse through WG1
and switch off a drive pulse through WG2.
(iii)~Reset stage. We simultaneously apply a classical reset pulse through WG1
and a classical drive pulse through WG2.
The carrier frequencies of the three pulses applied through WG1 
are different from each other.}
\label{fig:pulse}
\end{figure}

\section{capture of signal photon}\label{sec:capture}
\begin{figure}[t]
\includegraphics[scale=1.0]{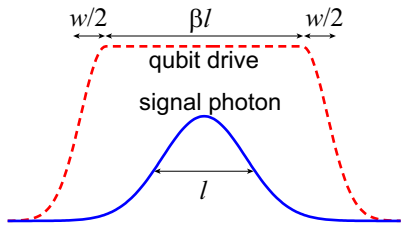}
\caption{
Envelopes of the signal photon and the qubit drive pulse. 
We use a smooth drive pulse in order to 
switch the I and $\Lambda$ modes adiabatically.
}
\label{fig:env}
\end{figure}

In this section, we present the numerical results 
for the capture stage of Fig.~\ref{fig:pulse},
where a signal pulse to be detected and a qubit drive pulse
are input simultaneously.
We assume a Gaussian pulse with length $l$ for the signal pulse:
\beq
f_s(t) = \left(\frac{8\ln 2}{\pi l^2}\right)^{1/4}
2^{-t^2/(l/2)^2}e^{-i\om_s t},
\eeq
Note that $f_s$ is normalized as $\int dt |f_s(t)|^2=1$.
On the other hand, we assume the following pulse shape for the qubit drive:
\beq
f_d(t)=\frac{\Om_d}{\sqrt{\gam^{\prime}}} e^{-i\om_d t}\times
\begin{cases}
1 & (|t| \leq \beta l/2) \\
2^{-(|t|-\beta l/2)^2/(w/2)^2} & (|t| > \beta l/2) 
\end{cases}.
\label{eq:fd}
\eeq
Namely, a square pulse with length $\beta l$ is smoothened 
by Gaussian functions with width $w$. 
We smoothen the drive pulses in order 
to switch the I and $\Lambda$ modes adiabatically and 
to suppress unwanted qubit excitations.
A real positive constant $\beta$ is chosen so that 
$f_d$ covers $f_s$ in time. 
For Gaussian signal pulses, we set $\beta=2$. 
The envelope functions of the signal photon and the drive pulse
are drawn in Fig.~\ref{fig:env}. 
In the capture stage, we set the drive amplitude 
at $\Om_d^{\rm imp}$ indicated in Fig.~\ref{fig:cr}(c).

\subsection{Zero-photon input}\label{ssec:dr_0}
\begin{figure}[t]
\includegraphics[scale=1.0]{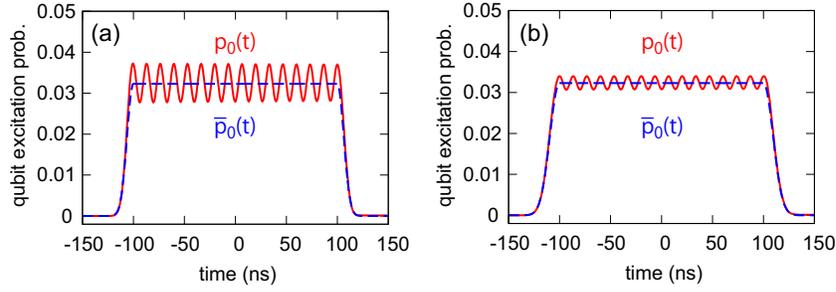}
\caption{
Time evolution of the qubit excitation probability, 
for a signal pulse containing zero photon: 
the actual evolution $p_0(t)$ (red solid)
and the adiabatic approximation $\overline{p}_0(t)$ (blue dashed).
The parameters used are: 
$\Om_d/2\pi=\Om_d^{\rm imp}/2\pi=13.2$~MHz, $l=100$~ns, $\beta=2$, 
and $w=20$~ns in (a) and $w=30$~ns in (b). 
}
\label{fig:dr0tev}
\end{figure}

Here, we discuss the case in which the signal pulse contains no photons.
Namely, we observe the dynamics induced solely by the drive pulse
applied through WG2.
In Fig.~\ref{fig:dr0tev}, we plot the time evolution 
of the qubit excitation probability, $p_0(t)=\la\sig^{\dag}\sig\ra$.
(The subscript $0$ represents the photon number in the signal pulse.) 
For reference, we also plot the excitation probability $\overline{p}_{0}(t)$
under the adiabatic approximation, which is obtained 
as follows. The system Hamiltonian including the qubit drive is written, 
in the rotating frame, as
$\cH_{\rm sys+dr} = \om_r a^{\dag}a \sig \sig^{\dag} +
\left[(\om_q-\om_d)+(\om_r-2\chi)a^{\dag}a \right] \sig^{\dag}\sig
+\sqrt{\gam'}|f_d(t)|(\sig^{\dag}+\sig)$.
Its lowest two eigenstates $|\tone\ra$ and $|\ttwo\ra$ are given by
$|\tone\ra=\cos\theta|g,0\ra-\sin\theta|e,0\ra$ and 
$|\ttwo\ra=\sin\theta|g,0\ra+\cos\theta|e,0\ra$, where
\beq
\theta(t)=\frac{1}{2}\arctan\left(\frac{2\sqrt{\gam^{\prime}} |f_d(t)|}{\om_q-\om_d}\right).
\eeq
Under the adiabatic assumption, the system 
always stays in its ground state, $|\tone\ra$.
The excitation probability, $\overline{p}_0(t)=|\la e,0|\tone\ra|^2$, 
is therefore given by
\beq
\overline{p}_0(t)=\sin^2\theta(t).
\eeq
We observe in Fig.~\ref{fig:dr0tev} that 
$p_0(t)$ is at most a few percents and 
roughly agrees with $\overline{p}_0(t)$. 
The deviation between them originates in the nonadiabatic transition,
which is suppressed by increasing $w$. 
Immediately after the drive is switched off, 
$p_0(t)$ comes back close to zero and becomes nearly independent of time. 
Hereafter, we set $w=30$~ns and define the detection probability 
$P_0$ as the qubit excitation probability at the end of this stage.
Namely, $P_0 = p_0(t_f)$, where we set $t_f=\beta l/2+50$~ns. 
Under this choice of $w$, the detection probability 
is negligibly small for the zero-photon input.
This means that the dark counts are suppressed 
nearly completely in the present scheme.
For example, $P_0 \simeq 10^{-4}$ for $l=200$~ns and $\beta=2$.

\subsection{One-photon input}\label{ssec:dr_1}
\begin{figure}[t]
\includegraphics[scale=1.0]{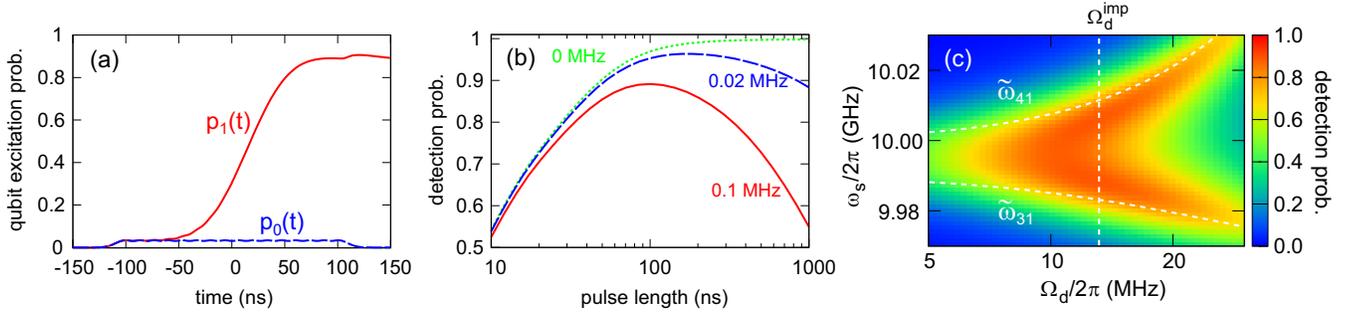}
\caption{
(a)~Time evolution of the qubit excitation probability $p_1(t)$ 
for a signal pulse containing one photon (red solid). 
The result for zero-photon case, $p_0(t)$, 
is also plotted for reference (blue dashed). 
The parameters used are: $l=100$~ns, $\beta=2$, 
$w=30$~ns, and $\om_s/2\pi=10.007$~GHz.
(b)~Dependence of the detection probability $P_1$
on the signal pulse length $l$ for various values of $\gam/2\pi$:
$0.1$~MHz (red solid), $0.02$~MHz (blue dashed) and $0$~MHz (green dotted).
(c)~Dependence of the detection probability $P_1$ 
on the drive amplitude $\Om_d$ 
and the signal photon frequency $\om_s$. 
}
\label{fig:dr1tev}
\end{figure}

Here, we discuss the case in which the signal pulse contains one photon.
We denote the qubit excitation probability for one-photon input by $p_1(t)$. 
We plot $p_1(t)$ in Fig.~\ref{fig:dr1tev}(a),
where the signal frequency is set at $\om_s/2\pi=10.007$~GHz
[the upper impedance-matching spot in Fig.~\ref{fig:cr}(d)]. 
It is observed that $p_1(t)$ increases rapidly 
during the signal pulse duration ($|t| \lesssim l/2$):
the excitation probability roughly evolves as 
$p_1(t)\sim\int^t d\tau|f_s(\tau)|^2$. 
At the end of the capture stage, 
a high detection probability $P_1=p_1(t_f)$ close to unity is achieved, 
in clear contrast with the zero-photon case where $P_0$ is negligibly small.

Figure~\ref{fig:dr1tev}(b) shows the dependence of $P_1$
on the signal pulse length $l$.
Here we assumed three values of the qubit decay rate $\gam$.
In the ideal limit of infinite qubit lifetime ($\gam=0$),
the detection probability increases monotonically in $l$
and a high probability close to unity is attained for $l \gg \kap^{-1}$.
This is because a longer pulse is advantageous for the destructive interference
between the input and elastically scattered photons,
which is the key physical mechanism for the
deterministic switching of the $\Lambda$ system.
However, in reality, the detection probability is decreased 
by the decay of the qubit into other channels during the pulse duration. 
A longer pulse is disadvantageous in this regard.
We thus have an optimum pulse length $l_{\rm opt}$ that maximize the detection probability. 
Under a practical qubit decay rate ($\gam/2\pi=$0.1~MHz),
$P_1$ amounts to 0.89 at $l_{\rm opt} \sim 100$~ns. 
If the qubit lifetime is improved five times ($\gam/2\pi=$0.02~MHz),
$P_1$ reaches 0.96 at $l_{\rm opt} \sim 200$~ns. 
Note that these detection probabilities represent
the switching probability in the capture stage of Fig.~\ref{fig:pulse}
which does not include the potential errors in the qubit readout stage.

Figure~\ref{fig:dr1tev}(c) plots $P_1$ 
as a function of the drive power and the signal frequency
for a signal pulse length $l=100$~ns.
We observe qualitative agreement with the reflectivity plot of Fig.~\ref{fig:cr}(d): 
$P_1$ is maximized when the drive amplitude satisfies 
the impedance-matching condition ($\Om_d=\Om_d^{\rm imp}$)
and the photon frequency is tuned closely to $\tom_{31}$ or $\tom_{41}$.

\subsection{Two-photon input}\label{ssec:dr_2}
\begin{figure}[t]
\includegraphics[scale=1.0]{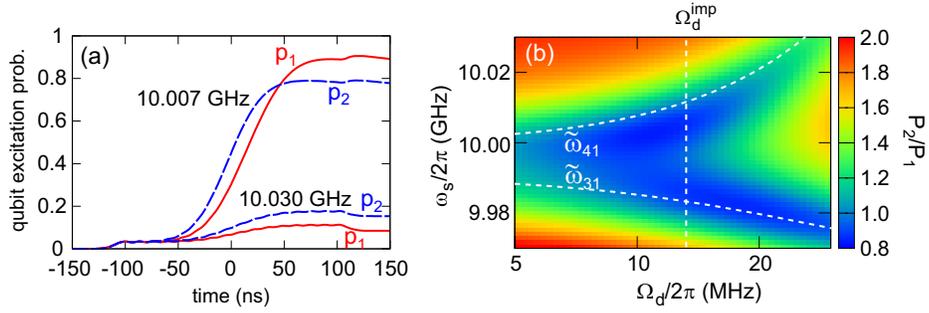}
\caption{
(a)~Time evolution of the qubit excitation probabilities, 
$p_1(t)$ (red solid) and $p_2(t)$ (blue dashed).
The input photons are on resonance 
with the $\Lambda$ system for $\om_s/2\pi=10.007$~GHz
and are out of resonance for $\om_s/2\pi=10.030$~GHz. 
(b)~Ratio of the detection probabilities, $P_2/P_1$,
as a function of $\Om_d$ and $\om_s$.
}
\label{fig:dr2tev}
\end{figure}

As we discussed in previous subsections,
our device discriminates with a high efficiency 
whether the signal pulse contains a photon or not.
Here we present the results for two-photon input 
to observe the multiphoton effects in the signal pulse.
In Fig.~\ref{fig:dr2tev}(a), 
we compare the time evolution of $p_{1}(t)$ and $p_{2}(t)$. 
When the signal photons are out of resonance with the $\Lambda$ system
($\om_s/2\pi=10.030$~GHz)
and therefore the excitation probabilities are low,
$p_{2}(t)$ is roughly twice as large as $p_{1}(t)$. 
The two photons excites the qubit independently in this case.
In contrast, when the signal photons are on resonance 
($\om_s/2\pi=10.007$~GHz)
and a high efficiency is attained for the one-photon input, 
$p_{2}(t)$ becomes comparable as $p_{1}(t)$
due to the saturation of the $\Lambda$ system. 
Figure~\ref{fig:dr2tev}(b) plots 
the ratio of the detection probabilities, $P_{2}/P_{1}$,
as a function of $\Om_d$ and $\om_s$.
By comparing Figs.~\ref{fig:dr1tev}(c) and \ref{fig:dr2tev}(b), we observe that 
$P_{2} \simeq 2P_{1}$ holds when $P_{1} \lesssim 0.2$ and 
$P_{2} \lesssim P_{1}$ holds when $P_{1} \gtrsim 0.7$.

\subsection{Pulse-shape dependence}\label{ssec:shape}
Up to here, we assumed a Gaussian pulse for the signal photon. 
Here, in order to observe the effects of the pulse shape, 
we consider square and exponential shapes for the signal pulse: 
\bea
f_{s}^{\rm sq}(t) &=& \frac{e^{-i\om_s t}}{\sqrt{l}} \times
\begin{cases}
1 & (|t| \leq l/2) \\
0 & (|t| > l/2) 
\end{cases},
\\
f_{s}^{\rm exp}(t) &=& \sqrt{\frac{2\ln 2}{l}}e^{-i\om_s t} \times
\begin{cases}
2^{-(t/l+\beta/2)} & (t > -\beta l/2) \\
0 & (t < -\beta l/2) 
\end{cases}.
\eea
Note that the amplitude decays exponentially in $f_{s}^{\rm exp}$,
as in the single photons generated by spontaneous emission of an emitter.
Regarding the drive pulse of Eq.~(\ref{eq:fd}), 
we set $\beta=1$ ($\beta=3$) for the square (exponential) shape
to cover the signal pulse.
We employ the same definition for the detection probability: 
$P_1=p_1(t_f)$, where $t_f=\beta l/2+50$~ns.

\begin{figure}[t]
\includegraphics[scale=1.0]{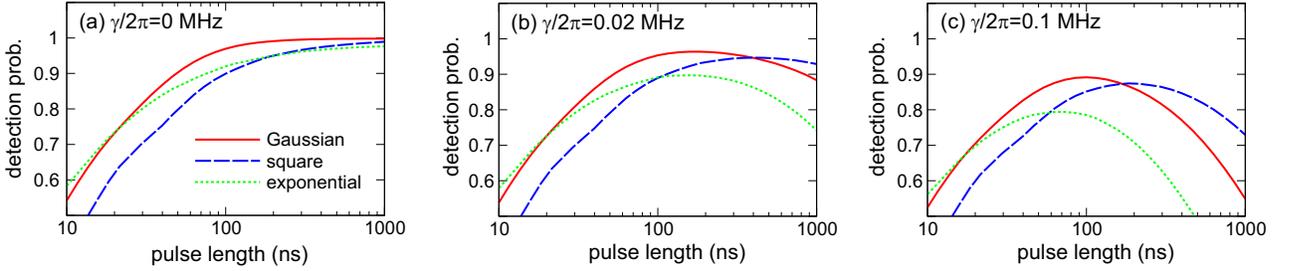}
\caption{
Dependence of the detection probability on the pulse length $l$
for various pulse shapes:
Gaussian (red solid), square (blue dashed), and exponential (green dotted).
The qubit decay rate $\gam/2\pi$ is $0$~MHz in (a),
$0.02$~MHz in (b), and $0.1$~MHz in (c).
}
\label{fig:ldep_s}
\end{figure}

Figure~\ref{fig:ldep_s} shows the dependence of 
the detection probability on the pulse length $l$. 
For the cases of infinite qubit lifetime [Fig.~\ref{fig:ldep_s}(a)], 
the detection efficiencies increase monotonically. 
A high detection probability exceeding 0.9 is obtained
for $l \gtrsim 100$~ns, regardless of the pulse shape.
However, if we take account of the finite qubit 
lifetime [Figs.~\ref{fig:ldep_s}(b) and (c)],
the detection probability decreases due to the qubit decay.
A square pulse is advantageous in this regard,
since we can use a shorter drive pulse (smaller $\beta$)
and suppress the qubit relaxation during the capture stage. 
Overall, regardless of the pulse shape, 
a high detection efficiency exceeding 0.9 is possible
if the qubit has a sufficient lifetime ($\gam/2\pi \lesssim 0.02$~MHz).

\section{reset of the system}\label{sec:reset}
In the present device, the qubit-resonator system is 
automatically reset to the ground state $|g,0\ra$ through the qubit decay.
However, in order to shorten the dead time of the detector,
we reset the system through microwave transitions.
In the reset stage of Fig.~\ref{fig:pulse},
we simultaneously apply a drive pulse to the qubit through WG2 
and a reset pulse to the resonator through WG1:
the drive pulse sets the system to the $\Lambda$ mode, 
and the reset pulse induces the $|\ttwo\ra \to |\widetilde{u}\ra \to |\tone\ra$ 
transition ($u=3$ or $4$).
The drive pulse profile is the same as that
in the capture stage, $f_d(t)$ of Eq.~(\ref{eq:fd}). 
However, as we observe below, a stronger drive 
($\Om_d>\Om_d^{\rm imp}$) is advantageous in this stage.
As the reset pulse, we use a classical Gaussian pulse, 
\beq
f_r(t) = \sqrt{\la n \ra}\times\left(\frac{8\ln 2}{\pi l^2}\right)^{1/4}
2^{-t^2/(l/2)^2}e^{-i\om_r t},
\eeq
where $\la n \ra$ is the mean photon number in the reset pulse.
Hereafter, we set $\om_r$ close to $\tom_{32}$
and use $|\ttwo\ra \to |\tthree\ra \to |\tone\ra$ transition for resetting.
Furthermore, we fix the pulse length at $l=100$~ns and $\beta=2$, 
and set the initial and final moments of the reset stage 
at $t_i=-\beta l/2-50$~ns and $t_f= \beta l/2+50$~ns.

\begin{figure}[t]
\includegraphics[scale=1.0]{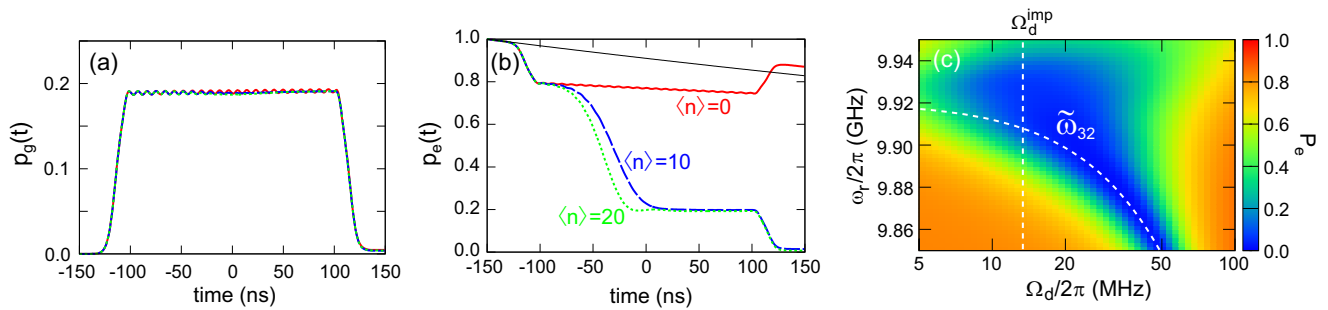}
\caption{
(a)~Time evolution of the qubit excitation probability 
$p_g(t)$ in the reset stage, starting from $|g,0\ra$.
The mean photon number in the reset pulse $\la n \ra$ is 
0 (red solid), 10 (blue dashed) and 20 (green dotted).
The three lines are mostly overlapping.
The parameters used are: 
$\Om_d/2\pi=44$~MHz, $\om_r/2\pi=9.860$~GHz,
$l=100$~ns, $\beta=2$, and $w=30$~ns. 
(b)~Time evolution of $p_e(t)$, starting from $|e,0\ra$.
The natural qubit decay is also plotted by the thin line.
(c)~Qubit excitation probability for $\la n \ra=10$
at the end of reset stage starting from $|e,0\ra$.
}
\label{fig:reset}
\end{figure}

In Fig.~\ref{fig:reset}(a), we show the time evolution
of the qubit excitation probability $p_g(t)$ in the reset stage, 
starting from $|g,0\ra$. 
For $t_i<t<-\beta l/2$, an adiabatic $|g,0\ra \to |\tone\ra$ transition
is induced by the drive pulse.
For $-\beta l/2 <t<\beta l/2$, 
the excitation probability is perturbed by the reset pulse only slightly,
which is due to the large detuning of $\om_r$
from both $\tom_{31}$ and $\tom_{41}$.
For $\beta l/2<t<t_f$, through an adiabatic $|\tone\ra\to|g,0\ra$ transition
the system returns to the ground state as the drive pulse is switched off.
At the end of the stage, the unwanted qubit excitation is suppressed nearly completely.
For example, the excitation probability $P_g=p_g(t_f)$
is below 0.005 for $\la n \ra \lesssim 20$.

In Fig.~\ref{fig:reset}(b), 
we show the qubit excitation probability $p_e(t)$ starting from $|e,0\ra$.
For $t_i<t<-\beta l/2$, an adiabatic $|e,0\ra \to |\ttwo\ra$ transition
is induced by the drive pulse.
For $-\beta l/2 <t<\beta l/2$, we observe rapid decrease of $p_e(t)$, 
which is due to the $|\ttwo\ra \to |\tthree\ra \to |\tone\ra$ transition 
induced by the reset pulse.
The reset pulse power $\la n \ra$ determines the transition rate.
When $\la n \ra \gtrsim 10$, the system transits 
to $|\tone\ra$ nearly completely at $t=\beta l/2$. 
For $\beta l/2<t<t_f$, through an adiabatic $|\tone\ra\to|g,0\ra$ transition
the system is reset to the ground state.
The excitation probability $P_e$ at the end of the stage 
is 0.015 (0.007) for $\la n \ra =10$ (20).
This is in clear contrast with the slow natural decay of the qubit,
where the excitation probability remains 0.83 at $t=t_f$.

In Fig.~\ref{fig:reset}(c), 
$P_e$ is plotted as a function of the drive amplitude $\Om_d$ and 
the reset pulse frequency $\om_r$. 
We observe that the impedance-matching condition ($\Om_d=\Om_d^{\rm imp}$)
is not necessarily required in the reset stage.
The system is reset efficiently when 
$\om_r \simeq \tom_{32}$ and $\Om_d > \Om_d^{\rm imp}$, 
which implies $\tkap_{31}>\tkap_{32}$ from Fig.~\ref{fig:cr}(a).
Although the reset efficiency per one reset photon 
is lowered by the unbalanced decay rate,
many photons ($\la n \ra \gtrsim 10$) involved in the pulse 
enables highly efficient resetting. 

\newpage
\section{summary}\label{sec:summary}
We theoretically analyzed the microwave response of 
a driven qubit-resonator system
and demonstrated its excellent performance 
as a detector of itinerant microwave photons. 
The detection of microwave photon proceeds 
by switching two modes adiabatically: 
Under a proper qubit drive,
the system functions as an impedance-matched $\Lambda$ system,
which efficiently absorbs itinerant photons ($\Lambda$ mode). 
Due to the destructive interference 
between the incident and elastically scattered photons,
a high efficiency close to unity is attained 
for temporally long pulses, regardless of their shapes.
In contrast, without a qubit drive, the system functions as an oscillator 
whose resonance frequency depends on the qubit state (I mode). 
We can perform dispersive readout of the qubit state in this mode.
The detector is operated in the following cycle:
In the capture stage, we set the system in the $\Lambda$ mode 
and map the presence (absence) of a photon 
in the signal pulse to the excited (ground) state of the qubit. 
In the readout stage, we set the system in the I mode 
and measure the qubit state. 
In the reset stage, in order to shorten the dead time of the detector,
we restore the system to its ground state through microwave transitions. 
The detection efficiency readily exceeds 0.9 for realistic parameters
and the dark counts are suppressed nearly completely.
The present scheme provides a long-sought single-photon 
detector in the microwave domain
and widens the options in microwave quantum-optics experiments.
It could be used for the characterization of photon number statistics
in nonclassical itinerant microwave fields
and the measurement of flying qubits encoded in propagating microwave photons.

\section*{Acknowledgment}
The authors are grateful to C. Eichler, E. Solano, 
and J. Taylor for fruitful discussion. 
This work was partly supported by 
MEXT KAKENHI (Grant Nos. 25400417 and 26220601),
Project for Developing Innovation Systems of MEXT, 
National Institute of Information and Communications Technology (NICT),
and ImPACT Program of Council for Science, Technology and Innovation.

\appendix
\section{Relation between coherent-state and Fock-state inputs}\label{app:A}
Here we derive Eq.~(\ref{eq:trans}), which connects the expectation values  
for a coherent-state input to those for a Fock-state input~\cite{KKPRL2007}.
We denote a coherent state by $|\alp\ra$ and 
a Fock state by $|m\ra$ ($m=0, 1, \cdots$). 
We denote the expectation value of an operator $S(t)$ 
for the coherent-state input by $\la S \ra_c = \la\alp|S|\alp\ra$
and its component proportional to $(\alp^*)^m\alp^n$ by $\la S \ra_c^{mn}$. 
On the other hand, we denote the matrix elements in the Fock bases
by $\la S \ra_{mn} = \la m|S|n\ra$.

Using the Fock-state expansion of a coherent state,
$|\alp\ra = e^{-|\alp|^2/2}\sum_{n=0}^\infty \alp^n|n\ra/\sqrt{n!}$,
$\la S \ra_c$ is written as
\beq
\la S \ra_c = e^{-|\alp|^2} \sum_{m,n} 
\frac{(\alp^*)^m\alp^n}{\sqrt{m!n!}} \la S \ra_{mn}.
\eeq
Picking up the terms proportional to $(\alp^*)^0\alp^0$, 
$(\alp^*)^1\alp^1$ and $(\alp^*)^2\alp^2$, we have
$\la S \ra_c^{00} = \la S \ra_{00}$, 
$\la S \ra_c^{11} = \la S \ra_{11}-\la S \ra_{00}$ and 
$\la S \ra_c^{22} = \la S \ra_{22}/2-\la S \ra_{11}+\la S \ra_{00}/2$. 
Therefore, 
\bea
\la S \ra_{00} &=& \la S \ra_c^{00},
\\
\la S \ra_{11} &=& \la S \ra_c^{00}+\la S \ra_c^{11},
\\
\la S \ra_{22} &=& \la S \ra_c^{00}+2\la S \ra_c^{11}+2\la S \ra_c^{22}.
\eea



\end{document}